\documentstyle[11pt]{article}

\begin{document}
\begin{center}
{\LARGE{\bf Inhomogeneous Bianchi type-I Cosmological Model with Electromagnetic Field in Lyra Geometry}}\\[1em]
\large{\bf{M. Abdel-Megied \footnote{E-mail : amegied@frcu.eun.eg}, Ragab M. Gad}\footnote{E-mail: ragab2gad@hotmail.com} and E. A. Hegazy\footnote{E-mail : sayedhegazy95@yahoo.com}}\\
\normalsize {Mathematics Department, Faculty of Science,}\\
\normalsize  {Minia University, 61915 El-Minia,  EGYPT.}
\end{center}

\begin{abstract}
We have investigated an inhomogeneous Bianchi type-I cosmological
model with electromagnetic field based on Lyra geometry. A new class
of exact solutions have been obtained by considering the potentials
of metric and displacement field are functions of coordinates $t$
and $x$. The physical behavior of the obtained model is discussed.
\end{abstract}

\setcounter{equation}{0}
\section{Introduction}
After Einstein \cite{E16} proposed his theory of general relativity,
which provided a geometrical description of gravitation, many
physicists attempted to generalize the idea of geometrizing the
gravitation to include a geometrical description of
electromagnetism. One of the first attempts was made by Weyl
\cite{W18} who proposed a more general theory, by formulating a new
kind of gauge theory involving metric tensor to geometrize
gravitation and electromagnetism. This theory was criticized due to
non-integrability of length of vector under parallel displacement
\cite{ R82, SD71}.
\par
 Later Lyra \cite{L51} suggested a modification
of Riemannian geometry by introducing a gauge function which removed
the non-integrability condition. This modified geometry known as
Lyra geometry. In a subsequent investigations Sen \cite{Sen57} and
Sen and Dunn \cite{SD71} formulated a new scalar-tensor theory of
gravitation and constructed an analog of the Einstein field
equations based on Lyra geometry.

 For physical motivation of
Lyra geometry we refer to the literature \cite{S60}-\cite{H73}.
 Sen \cite{Sen57}
found that static model with finite density in Lyra geometry is
similar to the static Einstein model, but a significant difference
that the model exhibited red shift. Halford \cite{H70} showed in his
study that the constant displacement vector field
 in Lyra geometry plays the role of  cosmological
constant in the theory of  general relativity. Halford \cite{H72}
showed that the scalar tensor treatment based in Lyra geometry
predicts the same effects, within observational limits, as in
Einstein theory.  Several attempts have been made to cast the scalar
tensor theory of gravitation in wider geometrical context \cite{S1}.
\par
Many Authors \cite{7} have studied cosmological models in Lyra
geometry with a constant gauge vector in the time-direction. The
close connection between these models and general relativistic
models has often been noted. Singh and his collaborators \cite{8}
have studied Bianchi type I, III, Kantowski-Sachs and new class of
models with a time dependent displacement field. They have made a
comparative study of Robertson-Walker models with a constant
deceleration parameter in Einstein's theory with a cosmological
terms and in the cosmological theory in Lyra geometry.
\par
Bianchi-type-I cosmological model has been studied by a number of
authors within the frame work of Lyra geometry. For instance, Bali
and Chandnani \cite{BC08} studied this model with time dependent
gauge function for perfect fluid distribution. Singh and Kale
\cite{SK09} investigated this model with uniform and variable
graviational constant with bulk viscosity. Spatially homogeneous
model, which is a transform form of Bianchi type-I space-time in
comoving coordinates, has been investigated, in the context of Lyra
geometry, for viscous fluid distribution by Pradhan \cite{P09} and
when the source of the gravitational field is perfect fluid
distribution by Reddy and Venkateswarlu \cite{RV88}. In this paper,
we investigate the evolution of Bianchi type-I cosmological model in
the presence of electromagnetic field within the frame work of Lyra
geometry.
\par
 In section 2, a review for the
basic equations of an anisotropic Bianchi type-I model in the
presence of electromagnetic field is given. In section 3, we
generate an exact  solution to the Einstein field equations. Section
4, deals with the study of some physical and geometrical properties
of the obtained model.

\setcounter{equation}{0}
\section{\bf{The metric and field equations}}
We consider Bianchi type-I metric in the form:
\begin{equation}\label{k-1}
ds^{2}=dt^{2}-A^{2}dx^{2}-B^{2}dy^{2}-C^{2}dz^{2},
\end{equation}
with the convention $(x^{0}=t$, $x^{1}=x$, $x^{2}=y$, $x^{3}=z)$,  $A$ is a function of $t$ only and $B$ and $C$ are functions of $x$ and $t$.\\

The volume element of the model (\ref{k-1}) is given by
\begin{equation}\label{vol}
V = \sqrt{-g}= ABC
\end{equation}
The four-acceleration vector, the rotation, the expansion scalar and
the shear scalar characterizing the four velocity vector field,
$u^i$, which satisfying the relation in co-moving coordinate system
$$
g_{ij}u^{i}u^{j}=1, \qquad u^{i}=u_{i}=(1,0,0,0),
$$
respectively, have the usual definitions as given by Raychaudhuri
\cite{R79}
\begin{equation}\label{Kin}
\begin{array}{ccc}
\dot{u}_i & = & u_{i;j}u^j ,\\
\omega_{ij} & = & u_{[i;j]}+\dot{u}_{[i}u_{j]},\\
\Theta & =  & u^i_{;i},\\
\sigma^2 & =& \frac{1}{2}\sigma_{ij}\sigma^{ij},
\end{array}
\end{equation}
where
$$
\sigma_{ij} =u_{(i;j)}+\dot{u}_{(i}u_{j)} -\frac{1}{3}\Theta
(g_{ij}+u_{i}u_{j}).
$$
In view of the metric (\ref{k-1}), the four-acceleration vector, the
rotation, the expansion scalar and the shear scalar given by
(\ref{Kin}) can be written in a comoving coordinates system as

\begin{equation}\label{Kin-v}
\begin{array}{ccc}
\dot{u}_{i} & =& 0,\\
\omega_{ij} & = & 0,\\
\Theta & = & (\frac{\dot{A}}{A}+\frac{\dot{B}}{B}+\frac{\dot{C}}{C}) ,\\
\sigma^2 & = & \frac{1}{3}\big(
\frac{\dot{A}^2}{A^2}+\frac{\dot{B^2}}{B^2}+
     \frac{\dot{C^2}}{C^2} - \frac{\dot{A}\dot{B}}{AB} - \frac{\dot{A}\dot{C}}{AC}- \frac{\dot{B}\dot{C}}{BC}\big),
\end{array}
\end{equation}
where the non vanishing components of the shear tensor
$\sigma_{i}^{j}$ are
\begin{equation}\label{comp}
\begin{array}{ccc}
\sigma_{1}^1 & =  &
\frac{1}{3}(\frac{2\dot{A}}{A}-\frac{\dot{B}}{B}-
     \frac{\dot{C}}{C}),\\
\sigma_{2}^2 & = & \frac{1}{3}(\frac{2\dot{B}}{B}-\frac{\dot{A}}{A}-
    \frac{\dot{C}}{C}),\\
 \sigma_{3}^3& = & \frac{1}{3}(\frac{2\dot{C}}{C}-\frac{\dot{A}}{A}-
    \frac{\dot{B}}{B}),\\
 \sigma_{4}^4& = 0 & .

\end{array}
\end{equation}
The field equations based in Lyra geometry as obtained by Sen
\cite{Sen57} can be written as:
\begin{equation}\label{k-2}
 G_{ij}+\frac{3}{2}\phi_{i}\phi_{j}-
 \frac{3}{4}g_{ij}\phi_{k}\phi^{k}=-\chi T_{ij},
\end{equation}
where $G_{ij}$ is the usual Einstein tensor, whereas $\phi_{i}$ is a
displacement field vector  defined by
\begin{equation}\label{k-3}
\phi_{i}=(\beta(x, t),0,0,0).
\end{equation}
$T_{ij}$ is the energy momentum tensor given by
 \begin{equation}\label{k-4}
 T_{ij}=(\rho+p)u_{i}u_{j}-pg_{ij}+E_{ij}.
\end{equation}
where $E_{ij}$ is the electro-magnetic field given by Lichnerowicz
\cite {li}:
\begin{equation}\label{}
    E_{ij}=\bar{\mu}[h_{l}h^{l}(u_{i}u_{j}-\frac{1}{2}g_{ij})+h_{i}h_{j}]
\end{equation}
Here  $\rho$ and $p$ are the energy density and isotropic pressure
respectively and $\bar{\mu}$ is the magnetic permeability and
$h_{i}$ the magnetic flux vector defined by:
\begin{equation}\label{}
    h_{i}=\frac{\sqrt{-g}}{2\bar{\mu}}\epsilon_{ijkl}F^{kl}u^{j}
\end{equation}
$F_{ij}$ is the electromagnetic field tensor and $\epsilon_{ijkl}$
is the Levi-Civita tensor density.
 If we consider that the current flow along z-axis, then $F_{12}$ is only non-vanishing component of $F_{ij}$.\\
The Maxwell's equations
\begin{equation}\label{}
    F_{ij;k}+F_{jk;i}+F_{ki;j}=0
\end{equation}
and \begin{equation}\label{}
    [\frac{1}{\bar{\mu}}F^{ij}]_{;j}=J^{i}
\end{equation}
require that $F_{12}$ be function of $x$ alone \cite {AP}. We assume
that the magnetic permeability as a function of $x$ and $t$ both.
Here the
semicolon represents a covariant differentiation.\\
For the line element (\ref{k-1}) the field equation (\ref{k-2}) with
equation (\ref{k-4}) lead to the following system of equations
\begin{equation}\label{k-7}
\frac{\dot{B}\dot{C}}{BC}+\frac{\ddot{B}}{B}+\frac{\ddot{C}}{C}-
\frac{1}{A^{2}}\frac{B^\prime
C^\prime}{BC}+\frac{3}{4}\beta^{2}=-\chi\big(p-\frac{F_{12}^{2}}{2\bar{\mu}
A^{2}B^{2}}\big),
\end{equation}
\begin{equation}\label{k-8}
\frac{\ddot{A}}{A}+\frac{\dot{A}\dot{C}}{AC}+\frac{\ddot{C}}{C}-
\frac{1}{A^{2}}\frac{C^{\prime\prime}}{C}+\frac{3}{4}\beta^{2}=
-\chi\big(p-\frac{F_{12}^{2}}{2\bar{\mu} A^{2}B^{2}}\big),
\end{equation}
\begin{equation}\label{k-9}
\frac{\ddot{A}}{A}+\frac{\dot{A}\dot{B}}{AB}+\frac{\ddot{B}}{B}-
\frac{1}{A^{2}}\frac{B^{\prime\prime}}{B}+\frac{3}{4}\beta^{2}=
-\chi\big(p+\frac{F_{12}^{2}}{2\bar{\mu} A^{2}B^{2}}\big),
\end{equation}
\begin{equation}\label{k-10}
  \frac{\dot{B}^{\prime}}{B}-\frac{\dot{A}B^{\prime}}{AB}+\frac{\dot{C}^{\prime}}{C}
  -\frac{\dot{A}C^{\prime}}{AC}=0,
\end{equation}

\begin{equation}\label{k-11}
\frac{\dot{A}\dot{B}}{AB}+\frac{\dot{A}\dot{C}}{AC}+\frac{\dot{B}\dot{C}}{BC}-
\frac{1}{A^{2}}\big(\frac{B^{\prime\prime}}{B}+\frac{C^{\prime\prime}}{C}+
\frac{C^{\prime}B^{\prime}}{CB}\big)-\frac{3}{4}\beta^{2}=
\chi\big(\rho-\frac{F_{12}^{2}}{2\bar{\mu} A^{2}B^{2}}\big),
\end{equation}
where  the over heat dot denotes differentiation with respect to $t$
and over head prime denotes differentiation with respect to $x$.
\par

\setcounter{equation}{0}
\section{Solutions of the Field Equations}

The field equations (\ref{k-7})-(\ref{k-11}) constitute a system of
 five highly non-linear differential  equations with seven unknowns variables, $A, B, C,  p,
\rho, F_{12}$ and $\beta$. Therefore, two physically reasonable conditions amongst these parameters are required to obtain explicit solutions of the field equations. \\

First, let us assume that the expansion scalar $\Theta$ in the model
(\ref{k-1}) is proportional to the eigenvalue $\sigma_{1}^{1}$ of
the shear tensor $\sigma_{i}^{j}$, then from (\ref{Kin-v}) and
(\ref{comp}) we get,

\begin{equation}\label{k-18}
    \frac{1}{3}(\frac{2\dot{A}}{A}-\frac{\dot{B}}{B}-\frac{\dot{C}}{C})=\alpha(\frac{\dot{A}}{A}+\frac{\dot{B}}{B}+\frac{\dot{C}}{C})
\end{equation}
where $\alpha$ is a constant of proportionality, hence;
\begin{equation}\label{k-19}
   \frac{\dot{A}}{A}=(\frac{3\alpha+1}{3})\frac{\dot{A}BC+\dot{B}AC+\dot{C}AB}{ABC},
\end{equation}
by integration we have:
\begin{equation}\label{k-20}
A=(BC)^{n},
\end{equation}
where \textbf{$n=\frac{3\alpha+1}{2-3\alpha}$} is a constant.\\

The motive behind assuming this condition is explained as follow
\cite{s-0}: Referring to Thorne \cite{Th} , the observations of
velocity-redshift relation for extra-galactic sources suggest that
the Hubble expansion of the universe is isotropic to within
$\approx$ 30 per cent\cite{s-1,s-2}. More precisely, the redshift
studies place the limit $\frac{\sigma}{H}$ $\leq 0.30$ where
$\sigma$ is the shear and $H$ the Hubble constant. Collins et al.
\cite{s-3} have pointed out that for a spatially homogeneous metric,
the normal congruence to the homogenous hypersurface satisfy the
condition $\frac{\sigma}{\Theta}$=constant.\\
 The second required condition is by assuming that the density $\rho$ and the pressure
$p$ are related by barotropic equation of state
 \begin{equation}\label{k-21}
    p=\lambda\rho \quad\quad\quad 0\leq\lambda\leq1.
 \end{equation}

From the condition (\ref{k-20}) we can make the following
assumptions

\begin{equation}\label{k-22}
   B(x,t)=f(x)k(t),\quad C(x,t)=\frac{l(t)}{f(x)}.
\end{equation}
Note that other choices can be made and thus the above definitions in equation (\ref{k-22}) are by no means unique.\\
If we substitute the assumptions in equation (\ref{k-22}) - take
into account the condition (\ref{k-20}) - into equation
(\ref{k-10}), we obtain the following relation for the functions
$k(t)$ and $l(t)$ as follows

\begin{equation}\label{k-23}
    k=c_{1}l,
\end{equation}
where $c_{1}$ is a constant of integration.\\
Using (\ref{k-20}), (\ref{k-22}) and (\ref{k-23}) in equations
(\ref{k-7}) and (\ref{k-8}), we get the following two equations
\begin{equation}\label{k-28}
    \frac{3f'^{2}}{f^{2}}-\frac{f''}{f}=a,
\end{equation}
where $a$ is a constant,
\begin{equation}\label{k-27}
    \frac{\ddot{l}}{l}+(2n+1)\frac{\dot{l}^{2}}{l^{2}}=ml^{-4n},
\end{equation}
 where $m=\frac{a}{(2n-1)c_{1}^{2n}}$ is a constant.\\

The solutions of the above two equations  (\ref{k-28}) and
(\ref{k-27}) are as follows
\begin{equation}\label{k-29}
    l(t)=(c_{3}t+c_{4})^{\frac{1}{2n}},
\end{equation}
\begin{equation}\label{k-30}
f(x)=(c_{5}e^{bx}+c_{6}e^{-bx})^{\frac{-1}{2}} ,
\end{equation}

where $c_{3}= n\sqrt{2m}, c_{4}=2n, b=\sqrt{2a}$, $c_{5}$ and
$c_{6}$ are
constants.\\
Consequently, equation  (\ref{k-23}) becomes
\begin{equation}\label{k-31}
    k(t)=c_{1}(c_{3}t+c_{4})^{\frac{1}{2n}}.
\end{equation}
 It is observed from equations (\ref{k-20}), (\ref{k-22}) and (\ref{k-29})-(\ref{k-31}) the metric potentials $A$, $B$ and $C$ can be singular only for $t \rightarrow \infty$. Thus the line element with these coefficients is singular free even at $t=0$, and can be written in the following form

    \[ds^{2}=dt^{2}-c_{1}^{2n}[c_{3}t+c_{4}]^{2}dx^{2}-c_{1}^{2}[c_{3}t
    +c_{4}]^{\frac{1}{n}}(c_{5}e^{bx}+c_{6}e^{-bx})^{-1}dy^{2}\]
\begin{equation}\label{33}
-[c_{3}t+c_{4}]^{\frac{1}{n}}(c_{5}e^{bx}+c_{6}e^{-bx})dz^{2}.
\end{equation}
The above line element can be transformed, by a proper choice of
constants, to the following line element
\begin{equation}\label{rm}
ds^2 = dt^2 -a_1^2(t) dx^2 - a_2^2(t) e^{2qx}dy^2 - a_3^2(t)
e^{-2qx}dz^2,
\end{equation}
where $q=\frac{b}{2}$. The line element (\ref{rm}) represents a
Bianchi type-$VI_o$.

\section{ Physical Properties of the Model}

 Using equations (\ref{k-22}) and (\ref{k-29})-(\ref{k-31}) in equations
 (\ref{k-7})-(\ref{k-11}), take into account the condition (\ref{k-20}), the expressions for density $\rho$, pressure $p$,  electromagnetic field $F_{12}$ and displacement field $\beta$ are given by
 \begin{equation}\label{k-36}
    \rho=\frac{c_{3}^{2}}{\chi(1-\lambda)n^{2}}T^{-2},
\end{equation}
\begin{equation}\label{k-37}
     p=\frac{\lambda c_{3}^{2}}{\chi(1-\lambda)n^{2}}T^{-2},
\end{equation}
\begin{equation}\label{k-35}
 F_{12}^{2}=\frac{\bar{\mu}
 c_{1}^{2(1+n)}}{\chi}\psi^{-1}T^{\frac{1}{n}}[\frac{(1-2n)c_{3}^{2}}{2n^{2}}
 +\frac{b^{2}}{c_{1}^{2n}}(\frac{\psi_{o}^{2}}{\psi^{2}}-\frac{1}{2})],
 \end{equation}
\begin{equation}\label{k-38} \frac{3}{2}\beta^{2}=[\frac{2\lambda c_{3}^{2}}{(\lambda-1)n^{2}}-2c_{3}^{2}-\frac{(1-4n^{2})c_{3}^{2}}{2n^{2}} +\frac{b^{2}}{2c_{1}^{2n}}\frac{\psi_{o}^{2}}{\psi^{2}}
]T^{-2},\end{equation}

 where \[\psi_{o}=c_{5}e^{bx}-c_{6}e^{-bx},\qquad
\psi=c_{5}e^{bx}+c_{6}e^{-bx}\]and \[T=c_{3}t+c_{4}\]
For the line element (\ref{33}), using equations (\ref{vol}), (\ref{Kin-v}) and (\ref{comp}), we have the following physical properties:\\
The volume element is
\begin{equation}\label{k-40}
    V=c_{1}^{n+1}T^{\frac{n+1}{n}}.
\end{equation}
This equation shows that the volume increases as the time increases, that is, the model (\ref{33}) is expanding with time.\\
The expansion scalar, which determines the volume behavior of the
fluid, is given by:
\begin{equation}\label{k-47}
    \Theta=\frac{c_{3}(n+1)}{nT}
\end{equation}
The  non-vanishing components of the shear tensor, $\sigma_{i}^{j}$,
are
\begin{equation}\label{k-42}
\sigma_{1}^{1}=\frac{(2n-1)c_{3}}{3nT},
\end{equation}
\begin{equation}\label{k-43}
\sigma_{2}^{2}=\frac{(1-2n)c_{3}}{6nT},
\end{equation}
\begin{equation}\label{k-44}
\sigma_{3}^{3}=\frac{(1-2n)c_3}{6nT},
\end{equation}
\begin{equation}\label{k-45}
\sigma_{4}^{4}=0.
\end{equation}
Hence, the shear scalar $\sigma$, is given by
\begin{equation}\label{k-46}
    \sigma=\frac{c_{3}(2n -1)}{2\sqrt{3}nT}
\end{equation}

Since $\lim_{t\rightarrow\infty} (\frac{\sigma}{\Theta})\neq 0$,
then the model (\ref{33}) does not approach isotropy for large value
of $t$. Also the model does not admit  acceleration and rotation,
since $\dot{u}_i =0$ and $\omega_{ij}=0$ .
We can see that
\begin{equation}\label{k-48}
    \frac{\sigma_{1}^{1}}{\Theta}=\frac{2n-1}{3(n+1)}=\alpha,
\end{equation}
which is a constant of proportional.
\section{Discussion and Conclusion}
We have offered an investigation of an anisotropic and an
inhomogeneous cosmological model of Bianchi type-I with
electromagnetic field in the context of Lyra geometry. We got an
interesting model which represents Bianchi type-$VI_o$.
 This model represents shearing and non-rotating. Moreover, this model is singular free even at the initial epoch  $t=0$ and has vanishing accelerations. We found also that $\lim_{t\rightarrow \infty}(\frac{\sigma}{\Theta}) \neq 0$, this means that the model does not approach isotropy for large time $t$.


\end{document}